\documentclass[useAMS,usenatbib]{mn2e}
\usepackage[dvipdfmx]{graphicx}
\usepackage{amsmath,amssymb,natbib,siunitx}
\usepackage{color}

\begin{document}

\title[Expected number of BBH detections by eLISA]{Concise estimate of
the expected number of detections for stellar-mass binary black holes by
eLISA}

\author[K. Kyutoku and N. Seto]
{Koutarou Kyutoku$^1$ and Naoki Seto$^2$\\
$^1$Interdisciplinary Theoretical Science (iTHES) Research Group, RIKEN,
Wako, Saitama 351-0198, Japan\\
$^2$Department of Physics, Kyoto University, Kyoto 606-8502, Japan}

\date{\today}

\maketitle

\begin{abstract}
 We study prospects for detecting extragalactic binary black holes
 similar to GW150914 by evolved Laser Interferometer Space Antenna
 (eLISA). We find that the majority of detected binary black holes will
 not merge within reasonable observation periods of eLISA in any
 configuration. While long-arm detectors are highly desired for
 promoting multiband gravitational-wave astronomy by increasing the
 detections of merging binaries, the number of total detections can be
 increased also by improving the acceleration noise. A monochromatic
 approximation works well to derive semiquantitative features of
 observational prospects for non-merging binaries with clearly
 indicating the parameter dependence. Our estimate also suggests that
 the number of galaxies in the error volume is so small that the host
 galaxy may be determined uniquely with high confidence.
\end{abstract}


\section{introduction} \label{sec:intro}

The first detection of a binary-black hole merger, GW150914, opened the
door to gravitational-wave astronomy \citep{ligovirgo2016}. The masses
of the black holes, $\sim 29$ and $\sim 36M_\odot$, are larger than
$\sim 10$--$15M_\odot$ previously expected from galactic observations
\citep{ozel_pnm2010,kreidberg_bfk2012}, and this finding prompts a
vigorous debate regarding their origin. At the same time,
\citet{ligovirgo2016-3} suggest a very high merger rate of binary black
holes in our Universe based on a part of Advanced LIGO's O1 observation
run. These facts immediately mean that space-based gravitational-wave
detectors, such as evolved Laser Interferometric Space Antenna
\citep[eLISA: see][for LISA Pathfinder]{lisapf2016}, would have a fair
chance to detect extragalactic stellar-mass binary black holes as well
as galactic compact binaries \citep{ligovirgo2016-2,sesana2016}.

Observing gravitational waves at low frequency will be important to
understand the origin of massive black holes like GW150914. Facing
massive black holes with $\gtrsim 30M_\odot$ unexpected for end products
of stellar evolution with the solar metallicity \citep{ligovirgo2016-2},
possible formation channels of the binary are actively discussed. One
plausible scenario is the evolution of low-metallicity stars with weak
stellar winds in an isolated field binary \citep[see][for
reviews]{postnov_yungelson2014}. Another scenario is dynamical formation
in dense stellar environments like galactic nuclei or globular clusters
\citep[see,
e.g.,][]{benacquista_downing2013,rodriguez_mpchr2015,rodriguez_hckr2016}. It
is also pointed out that a binary of primordial black holes satisfying
current observational constraints is also consistent with GW150914
\citep{bird_cmakkrr2016,sasaki_sty2016}. Because these scenarios predict
different distribution of binary parameters such as the eccentricity,
which is determined to higher accuracy at lower frequency \citep[see
section 1 of][for the discussion]{nishizawa_bks2016}, multiple
detections could statistically clarify the formation scenario \citep[see
also][]{breivik_rlkr2016,nishizawa_bks2016-2}. Moreover, precise
localization of the binary by the annual modulation of the detector with
the distance estimation will be beneficial to determine the host galaxy,
information of which is also invaluable to infer the origin.

In this paper, we study prospects for detecting extragalactic binary
black holes by eLISA, enhancing the previous investigation for Galactic
binary black holes by one of the authors \citep{seto2016}. The
possibility of detecting extragalactic binary black holes is mentioned
briefly by LIGO Scientific Collaboration immediately after the detection
of GW150914 \citep{ligovirgo2016-2}. Various authors conducted follow-up
studies of this possibility by Monte Carlo simulations
\citep{nishizawa_bks2016,sesana2016,vitale2016}, primarily focusing on
the aspects of multiband gravitational-wave astronomy, i.e.,
simultaneous detections of the same binary by eLISA and ground-based
interferometric detectors such as Advanced LIGO. Here, we analytically
evaluate the expected number of detections and put more emphasis on
extragalactic binary black holes that do not merge during the operation
of eLISA than previous studies do, because such binaries inevitably
dominate the detection. We also show that a monochromatic approximation
works well to derive semiquantitative features of observational
prospects for non-merging binaries.

Our assumptions and parameter choices are summarized as follows. We
treat all the binary black holes as circular, and denote the
gravitational-wave frequency by $f$, which is twice the orbital
frequency. This is justified to the accuracy of our discussion, because
the eccentricity is not expected to be very high. We apply the
quadrupole formula for point masses neglecting the black hole spin as
well as higher order post-Newtonian effects, and the cosmological
redshift is also neglected. We take the fiducial chirp mass of binary
black holes to be $\mathcal{M} = 28 M_\odot$ according to the estimate
from GW150914 \citep{ligovirgo2016}, and later show that our estimate
applies approximately to distribution of chirp masses once averaged over
the weight $\mathcal{M}^{10/3}$. We take the fiducial comoving merger
rate to be $R = \SI{100}{Gpc^{-3}.yr^{-1}}$ motivated by the estimate
from a part of O1 \citep{ligovirgo2016-3}.

\section{search for extragalactic binary black holes}

In this section, we derive the frequency distribution of the expected
number of detections for extragalactic binary black holes by eLISA with
various planned sensitivities. We take the fiducial value of the
observation period $T$ to be \SI{3}{yr} in the same manner as
\citet{seto2016} and the fiducial value of the detection threshold for
the signal-to-noise ratio $\rho_\mathrm{thr}$ to be 8.

\subsection{formulation}

The signal-to-noise ratio $\rho$ for a binary with the initial frequency
$f_i$ at the start of the observation is given by
\begin{equation}
 \rho^2 = 4 \int_{f_i}^{f_f} \frac{| \tilde{h} (f) |^2}{(3/20)S(f)} df ,
  \label{eq:snr}
\end{equation}
where $S(f)$ is the one-sided, sky-averaged noise spectral density of
eLISA such as those provided in \citet{elisa2012,klein_etal2016}. The
factor of $3/20$ is introduced to derive an effective non-sky-averaged
noise spectral density \citep{berti_bw2005,nishizawa_bks2016}. The
frequency $f_f$ at the end of the observation is given by
\begin{equation}
 f_f ( f_i , T ) = \left( f_i^{-8/3} - \frac{256 \pi^{8/3} G^{5/3}
                    \mathcal{M}^{5/3} T}{5 c^5} \right)^{-3/8} ,
\end{equation}
as long as the binary does not merge within the observation period. The
frequency above which the merger occurs within the observation period
$T$ is given by
\begin{align}
 f_\mathrm{merge} (T) & = \frac{5^{3/8} c^{15/8}}{8 \pi G^{5/8}
 \mathcal{M}^{5/8} T^{3/8}} \\
 & = \SI{19.2}{\milli\hertz} \left( \frac{\mathcal{M}}{28M_\odot}
 \right)^{-5/8} \left( \frac{T}{\SI{3}{yr}} \right)^{-3/8} ,
 \label{eq:fmerge}
\end{align}
and such binaries will serve as interesting targets of multiband
gravitational-wave astronomy
\citep{sesana2016,vitale2016,nishizawa_bks2016}. If the binary merges
within the observation period, namely $f_i \ge f_\mathrm{merge}(T)$, or
$f_f$ becomes higher than \SI{1}{\hertz}, above which the detector noise
would lose control, we set $f_f$ in equation \eqref{eq:snr} to be
\SI{1}{\hertz}.

The gravitational-wave spectral amplitude for a binary at a distance $D$
is given by
\begin{equation}
 \left| \tilde{h}(f) \right|^2 = \frac{5 G^{5/3} \mathcal{M}^{5/3}}{24
  \pi^{4/3} c^3 D^2 f^{7/3}} \times \frac{3}{4} \left[ F_+^2 \left(
                                                              \frac{1 +
                                                              \mu^2}{2}
                                                             \right)^2 +
  F_\times^2 \mu^2 \right] ,
\end{equation}
where $F_+$ and $F_\times$ are the antenna pattern functions, which
depend on the sky location and polarization angle, and $\mu = \cos
\imath$ is the cosine of the inclination angle $\imath$. We explicitly
separate the factor $(\sqrt{3}/2)^2$ accounting the opening angle
\ang{60} of eLISA from the antenna pattern functions. The antenna
pattern functions also change with time due to the annual rotation of
the detector plane.

Because we focus on the case that $T \ge \SI{1}{yr}$, we take the time
average of the antenna pattern functions in advance to set $\langle
F_+^2 \rangle_t = \langle F_\times^2 \rangle_t = 1/5$ for individual
binaries. This is a reasonable approximation at least for a six-link
configuration \citep{seto2004}.\footnote{For a binary with a short
signal duration, the average over the sky location is less than
0.221. Given the convex nature of the function $x^{3/2}$ that relates
the signal-to-noise ratio and the effective volume, the actual value
should be in the range $[0.2,0.221]$, resulting in an error less than
17\% for our effective volume below.} Meanwhile, even for a four-link
configuration, we can extend the procedure developed in \citet{seto2014}
to show that the averaging with respect to the polarization angle has a
negligible effect of $\sim 1\%$ level on the detectable volume $V ( f_i
, T )$ estimated below. We assume a four-link configuration of eLISA in
this study, and the result for a six-link configuration may be estimated
simply multiplying $\tilde{h}(f)$ by a factor of $\sqrt{2}$ and scaling
all the relevant quantities accordingly.

Then, the signal-to-noise ratio relevant for our study is expressed as
\begin{align}
 \rho^2 & = \frac{A^2}{D^2} \left[ \left( \frac{1 + \mu^2}{2} \right)^2
 + \mu^2 \right] I_7 ( f_i , T ), \label{eq:snrav} \\
 A^2 & \equiv \frac{G^{5/3} \mathcal{M}^{5/3}}{8 \pi^{4/3} c^3} ,
 \label{eq:snrfac} \\
 I_7 ( f_i , T ) & \equiv \int_{f_i}^{f_f} \frac{f^{-7/3}}{(3/20)S(f)}
 df .
\end{align}
Note that $I_7$ defined above depends on the initial frequency $f_i$ and
the observation period $T$. The maximum detectable distance within which
the signal-to-noise ratio is larger than a given threshold
$\rho_\mathrm{thr}$ depends on the inclination angle, initial frequency,
and observation period as
\begin{equation}
 D_\mathrm{thr} = \frac{A}{\rho_\mathrm{thr}} \left[ \left( \frac{1 +
                                                      \mu^2}{2}
                                                     \right)^2 + \mu^2
                                              \right]^{1/2} I_7 ( f_i ,
 T )^{1/2} .
\end{equation}
The detectable volume averaged over the inclination angle becomes a
function of the initial frequency and observation period as
\begin{equation}
 V ( f_i , T ) = \frac{4\pi}{3} \times \frac{1}{2} \int_{-1}^1
  D_\mathrm{thr}^3 d \mu ,
\end{equation}
and the effective range (or sensemon range) of eLISA is also defined by
$D_\mathrm{eff} \equiv ( 3V / 4\pi )^{1/3}$. Using \citep[see
also][]{schutz2011}
\begin{equation}
 \frac{1}{2} \int_{-1}^1 \left[ \left( \frac{1 + \mu^2}{2} \right)^2 +
                          \mu^2 \right]^{3/2} d \mu \approx 0.822 ,
 \label{eq:ave}
\end{equation}
the detectable volume averaged over the inclination angle is calculated
as
\begin{equation}
 V ( f_i , T ) = \frac{4 \pi}{3} \times 0.822
  \frac{A^3}{\rho_\mathrm{thr}^3} I_7 ( f_i , T )^{3/2} ,
  \label{eq:volume}
\end{equation}
and the effective range is given by
\begin{equation}
 D_\mathrm{eff} ( f_i , T ) = 0.937 \frac{A}{\rho_\mathrm{thr}} I_7 (
  f_i , T )^{1/2} . \label{eq:range}
\end{equation}
Hereafter, we denote the initial frequency simply by $f$ anticipating no
confusion would arise.

The frequency distribution of the expected number of detections for
binary black holes can be calculated as
\begin{equation}
 \frac{dN}{d \ln f} = V (f,T) \frac{dn}{d \ln f} , \label{eq:exact}
\end{equation}
where $dn/d \ln f$ is the number-density distribution of the
binaries. For a collection of identical binary black holes, the
distribution of the number density $n$ in each logarithmic frequency
interval should be proportional to the time that binaries spend in the
interval. Here, we specifically consider the collection of binary black
holes similar to GW150914. Using the comoving merger rate $R = dn/dt$,
the distribution is written as
\begin{align}
 \frac{dn}{d \ln f} & = \frac{f}{\dot{f}} R = \frac{5 c^5 R}{96
 \pi^{8/3} G^{5/3} \mathcal{M}^{5/3} f^{8/3}} \label{eq:distrib} \\
 & = \SI{4.57e-6}{Mpc^{-3}} \notag \\
 & \times \left( \frac{f}{\SI{10}{\milli\hertz}} \right)^{-8/3} \left(
 \frac{\mathcal{M}}{28M_\odot} \right)^{-5/3} \left(
 \frac{R}{\SI{100}{Gpc^{-3}.yr^{-1}}} \right) .
\end{align}
The distribution of the expected number of detections, $dN/d \ln f$, is
obtained by multiplying equations \eqref{eq:volume} and
\eqref{eq:distrib}, where it is proportional to $R /
\rho_\mathrm{thr}^3$ as expected. The distance $D_\mathrm{near}$ to the
nearest and thus loudest binary black holes in each logarithmic
frequency interval can be guessed from this number-density distribution
via the condition
\begin{equation}
 \frac{4\pi}{3} D_\mathrm{near}^3(f) \frac{dn}{d \ln f} = 1 ,
\end{equation}
and it is found to be
\begin{align}
 D_\mathrm{near} (f) & = \frac{3^{2/3} 2 \pi^{5/9} G^{5/9}
 \mathcal{M}^{5/9} f^{8/9}}{5^{1/3} c^{5/3} R^{1/3}} \\
 & = \SI{37.4}{Mpc} \notag \\
 & \times \left( \frac{f}{\SI{10}{\milli\hertz}} \right)^{8/9} \left(
 \frac{\mathcal{M}}{28 M_\odot} \right)^{5/9} \left(
 \frac{R}{\SI{100}{Gpc^{-3}.yr^{-1}}} \right)^{-1/3} .
\end{align}

\begin{figure}
 \includegraphics[width=.95\linewidth]{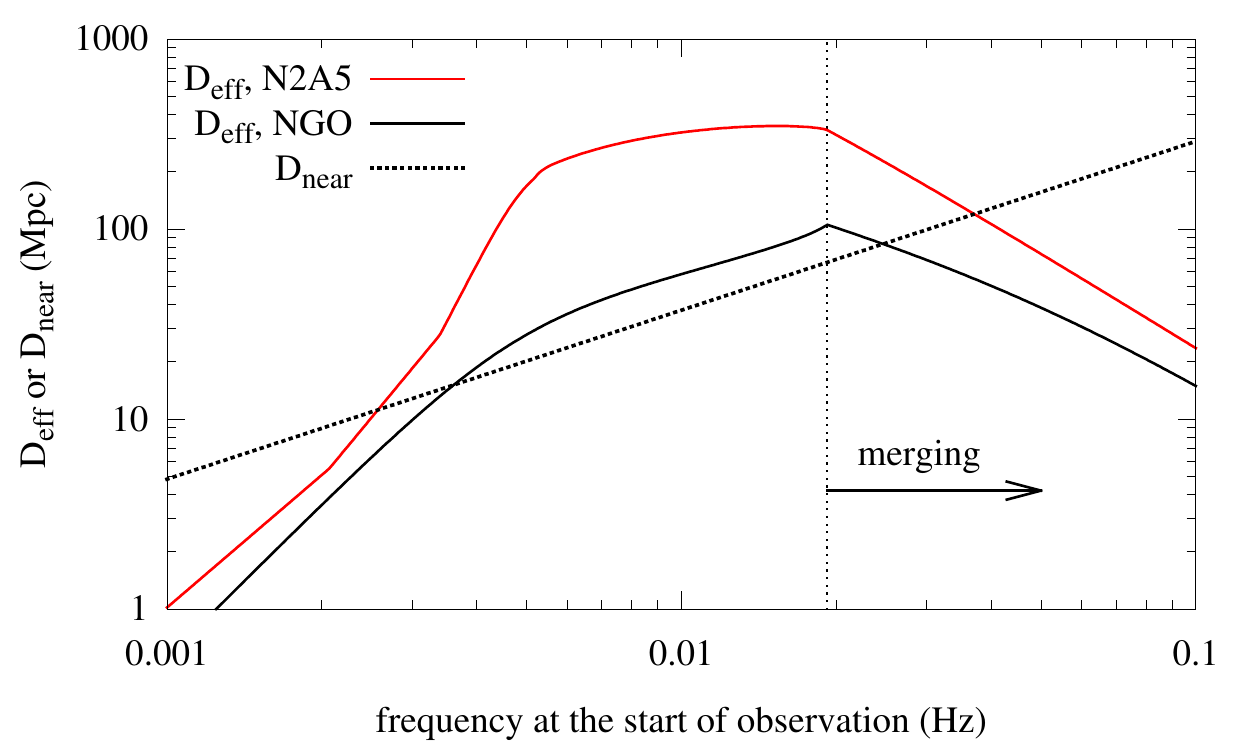} \caption{Effective
 range $D_\mathrm{eff}$ for 3-yr observations of eLISA and the distance
 $D_\mathrm{near}$ to the nearest binary black holes similar to
 GW150914. The curves for $D_\mathrm{eff}$ labelled by NGO and N2A5 are
 calculated with the noise curve of \citet{elisa2012} and that of
 \citet{klein_etal2016}, respectively. Galactic binary white dwarfs are
 taken into account as the foreground for the latter according to
 \citet{klein_etal2016}. The vertical dotted line marks
 $f_\mathrm{merge} ( \SI{3}{yr} ) = \SI{19.2}{\milli\hertz}$.}
 \label{fig:range}
\end{figure}

To indicate the relevant distance scale of the problem, we show the
effective range $D_\mathrm{eff} (f)$ for two representative noise curve
models of eLISA \citep{elisa2012,klein_etal2016} and the distance to the
nearest binary $D_\mathrm{near} (f)$ in Fig.~\ref{fig:range}. This
figure shows that the relevant range is local with $\lesssim
\SI{350}{Mpc}$. The expected number of detections becomes lower than
unity for the frequency range where $D_\mathrm{eff} \lesssim
D_\mathrm{near}$, and thus the actual number will fluctuate
significantly. This fluctuation will especially be relevant for the
detection of merging binary black holes satisfying $f > f_\mathrm{merge}
(T)$ by low-sensitivity configurations such as NGO \citep{elisa2012},
with which $D_\mathrm{eff} / D_\mathrm{near}$ never exceeds 2
irrespective of the frequency.

\subsection{Frequency distribution of the expected number of detections}

\begin{figure}
 \includegraphics[width=.95\linewidth]{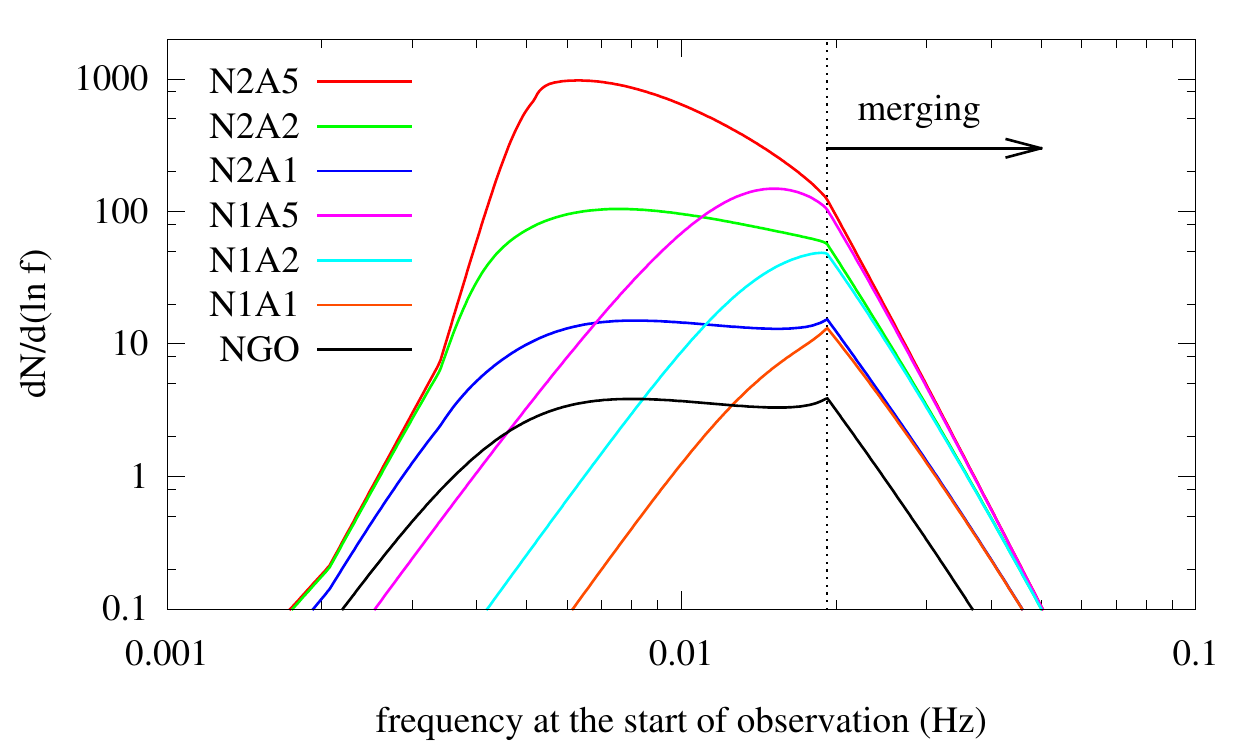} \caption{Frequency
 distribution of the expected number of detections for binary black
 holes similar to GW150914 in each logarithmic frequency interval for
 3-yr observations of eLISA. The curve labeled by NGO is calculated with
 the noise curve of \citet{elisa2012}, and the others are with those of
 \citet{klein_etal2016}. Galactic binary white dwarfs are taken into
 account as the foreground for N2 configurations according to
 \citet{klein_etal2016}. The vertical dotted line marks
 $f_\mathrm{merge} ( \SI{3}{yr} ) = \SI{19.2}{\milli\hertz}$.}
 \label{fig:dndlnf}
\end{figure}

Fig.~\ref{fig:dndlnf} shows the distribution, $dN/d \ln f$ (equation
\ref{eq:exact}), calculated with various noise curve models of eLISA
\citep{elisa2012,klein_etal2016}. For models of \cite{klein_etal2016},
N2 has a weaker acceleration noise than N1, and the sensitivity is
improved primarily at low frequency. A1, A2, and A5 correspond to the
arm length of \SI{1e6}{\kilo\meter}, \SI{2e6}{\kilo\meter}, and
\SI{5e6}{\kilo\meter}, respectively, and the long arm length improves
the sensitivity at intermediate frequency.

This figure shows that the noise curve has a critical impact on the
number of detections and also that the majority of the detected binaries
will not merge within the observation period in any case. The number of
merging binary black holes satisfying $f \ge f_\mathrm{merge} (T)$ is
not affected significantly by the sensitivity at low frequency, as found
from the comparisons between N1 and N2 configurations. This is expected,
because additional detections of merging binary black holes require
improvement of the sensitivity at $f \ge f_\mathrm{merge} (T)$. On
another front, a long arm length increases the detections of merging
binary black holes significantly, as found from the comparisons among
A1, A2, and A5 configurations. Both improvements enhance the number of
total detections.

\begin{figure}
 \includegraphics[width=.95\linewidth]{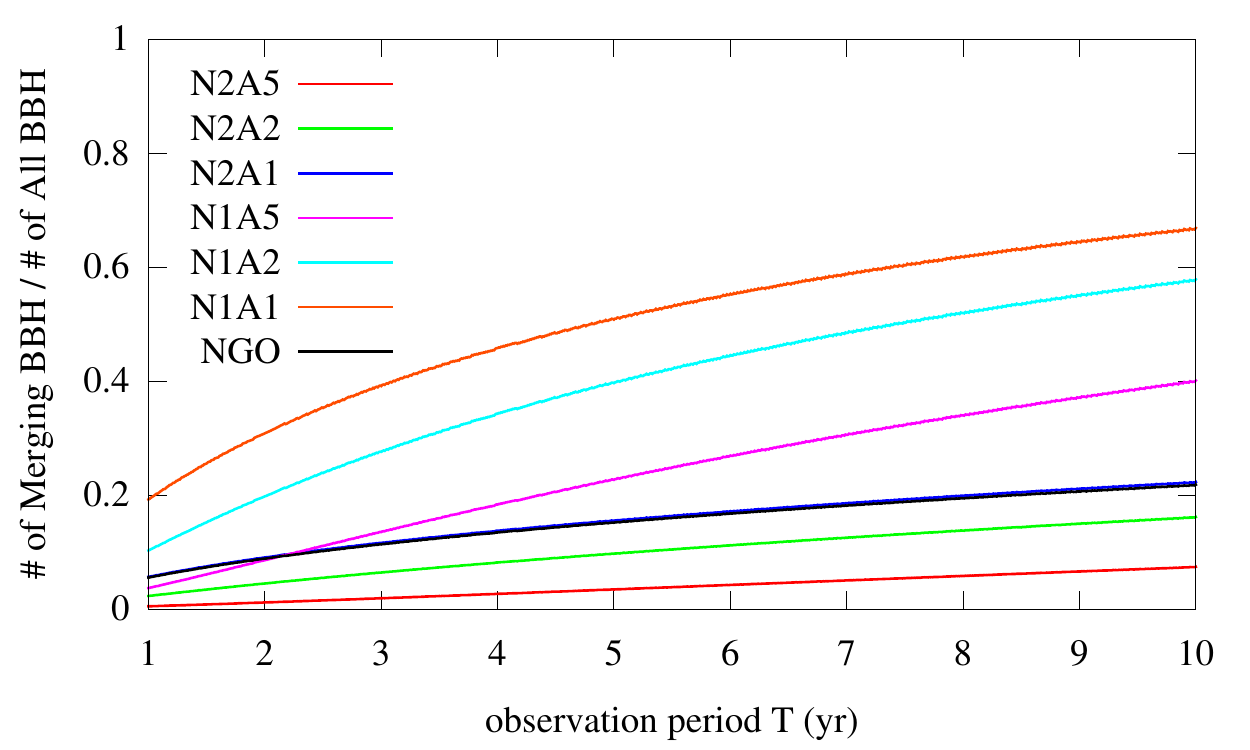} \caption{Fraction of
 detected binary black holes that merge within the observation period of
 eLISA to all the detected binary black holes as a function of the
 observation period. The curve for N2A1 mostly overlaps with that for
 NGO due to the similar shape of the noise curve.} \label{fig:fraction}
\end{figure}

The dominance of non-merging binary black holes is always the case for
reasonable observation periods as shown in Fig.~\ref{fig:fraction}. This
figure shows that merging binary black holes become dominant only for an
improbably long observational period of $T \gtrsim \SI{5}{yr}$ with the
N1 configurations, which are less sensitive at low frequency \citep[see
also][]{sesana2016}. The fraction will be no larger than $\sim 10\%$ for
the N2 configurations because of its high sensitivity at low frequency,
which drastically increases the detections of non-merging binary black
holes. The fraction is nearly the same for NGO and N2A1, because they
have very similar noise curves except for the overall normalization.

\begin{figure*}
 \begin{tabular}{cc}
  \includegraphics[width=.47\linewidth]{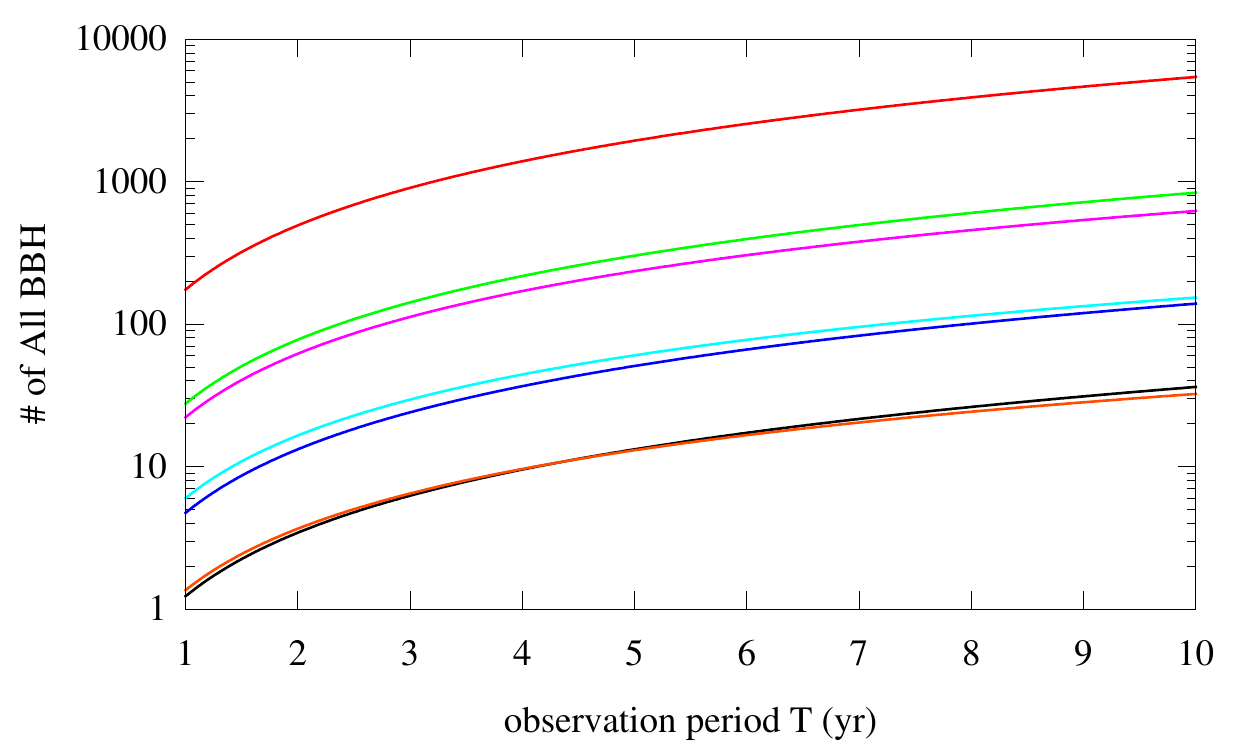} &
  \includegraphics[width=.47\linewidth]{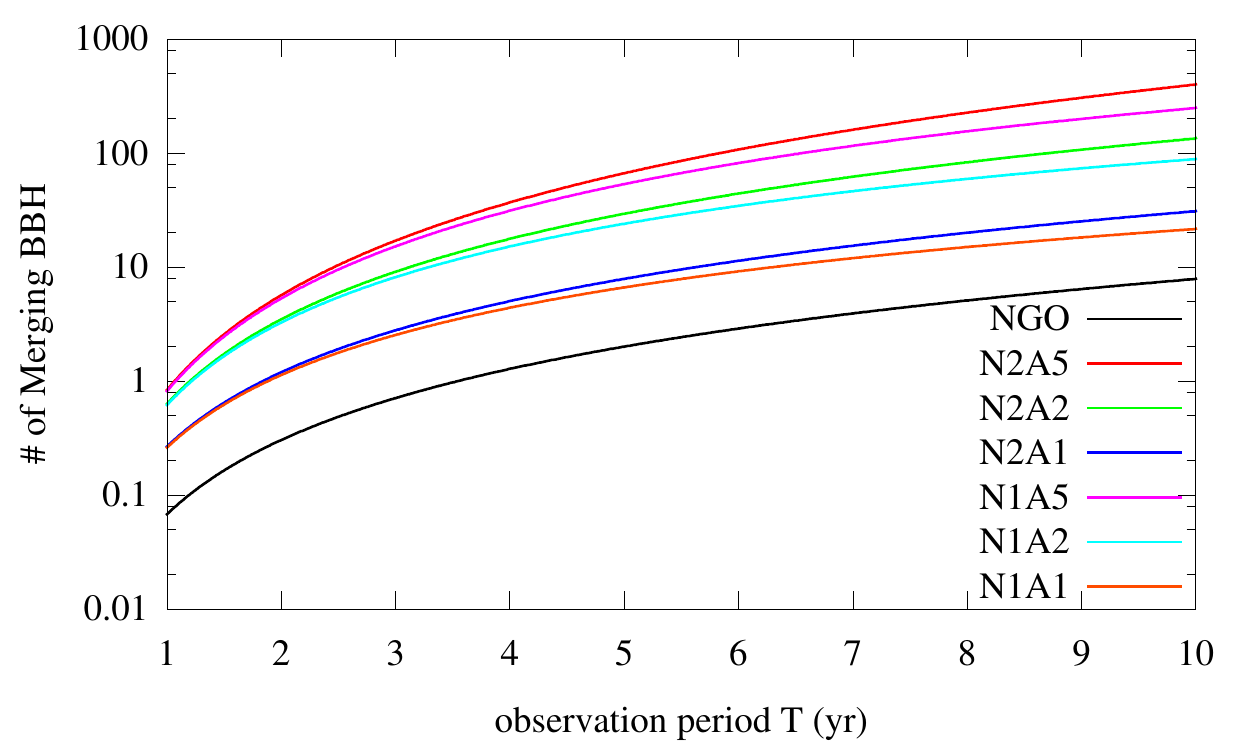}
 \end{tabular}
 \caption{Expected number of all the detected binary black holes (left)
 and merging binary black holes (right) as functions of the observation
 period of eLISA. Note the different scale of the vertical axes.}
 \label{fig:number}
\end{figure*}

\begin{table}
 \caption{The expected number of all the detected binary black holes and
 merging binary black holes for representative values of the observation
 period, $T$.}
 \centering
 \begin{tabular}{c|cccccc} \hline
  Model & \SI{1}{yr} & \SI{2}{yr} & \SI{3}{yr} & \SI{4}{yr} & \SI{5}{yr}
  & \SI{10}{yr} \\
  \hline \hline
  NGO & 1.2 & 3.4 & 6.2 & 9.5 & 13.2 & 36.2 \\
  (merge) & 0.07 & 0.3 & 0.7 & 1.3 & 2.0 & 7.9 \\
  N1A1 & 1.4 & 3.7 & 6.5 & 9.6 & 13.0 & 32.3 \\
  (merge) & 0.3 & 1.1 & 2.5 & 4.4 & 6.6 & 21.6 \\
  N1A2 & 6.0 & 16.5 & 29.5 & 44.3 & 60.3 & 153 \\
  (merge) & 0.6 & 3.3 & 8.1 & 15.2 & 23.9 & 88.8 \\
  N1A5 & 22.2 & 61.9 & 112 & 170 & 235 & 621 \\
  (merge) & 0.8 & 5.3 & 15.1 & 31.2 & 53.2 & 249 \\
  N2A1 & 4.7 & 13.2 & 24.0 & 36.6 & 50.8 & 139 \\
  (merge) & 0.3 & 1.2 & 2.8 & 5.0 & 7.9 & 31.0 \\
  N2A2 & 27.6 & 77.6 & 142 & 217 & 302 & 835 \\
  (merge) & 0.6 & 3.5 & 9.1 & 17.7 & 29.2 & 135 \\
  N2A5 & 174 & 492 & 903 & 1390 & 1940 & 5420 \\
  (merge) & 0.8 & 5.7 & 17.0 & 36.9 & 66.3 & 401 \\
  \hline
 \end{tabular}
 \label{table:number}
\end{table}

The number of merging binary black holes itself is always larger for the
N2 configurations than the N1 configurations as well as that of all the
detected binary black holes as shown in Fig.~\ref{fig:number}. We also
present the numbers for representative values of $T$ in Table
\ref{table:number}. This figure shows that the number of detections for
extragalactic stellar-mass binary black holes could be comparable to
that for supermassive binary black holes \citep{klein_etal2016}. This
figure also shows that the number of (merging or total) detected
binaries increases faster than the increase of the observation period,
$T$. As we explain in the next section, the total number is expected to
increase approximately as $T^{3/2}$ for the case that dominant sources
do not evolve significantly, i.e., the monochromatic approximation
applies well. This is the case for eLISA. This situation should be
contrasted with compact binary coalescences for ground-based detectors,
the number of which should be proportional to $T$. In reality, the
increase could be even faster for eLISA, because the long-term
observation will allow us to remove more confusion noise caused by
Galactic compact binaries. The total expected number of detections
varies by two orders of magnitude, from $\sim 6$ (NGO) to $\sim 900$
(N2A5) for $T=\SI{3}{yr}$, among different detector configurations.

The prospect for multiband gravitational-wave astronomy crucially
depends on the detector arm length. If the arm length is $\sim
\SI{1e6}{\kilo\meter}$ (A1), the number of binaries that merge within
the observation period is $O(1)$ at best. Even if we would count
binaries that merge within $\sim \SI{10}{yr}$ after the shutdown of
eLISA, the number will be only $\lesssim 30$. By contrast, the arm
length of $\sim \SI{5e6}{\kilo\meter}$ (A5) would increase the number of
merging binary black holes by a factor of several. Thus, the long arm
length of eLISA is highly desired for fruitful multiband
gravitational-wave astronomy. Reducing the acceleration noise adds not
much to multiband observations. Our estimates agree approximately with
those of \citet{sesana2016} derived by Monte Carlo simulations with
distribution of chirp masses.

\section{monochromatic approximation}

In the low-frequency regime where the binary evolution is negligible
during the observation period, the monochromatic approximation works
well to estimate various aspects of detectable signals. The
signal-to-noise ratio of gravitational waves is approximated by
\begin{align}
 \rho^2 & \approx 4 \frac{| \tilde{h} (f) |^2}{(3/20)S(f)} \dot{f} T \\
& = \frac{12 \pi^{4/3} G^{10/3} \mathcal{M}^{10/3}}{5 c^8 D^2} \left[
 \left( \frac{1 + \mu^2}{2} \right)^2 + \mu^2 \right]\frac{f^{4/3}
 T}{(3/20)S(f)} ,
\end{align}
where the time and polarization angle average are assumed. After
averaging over the inclination angle, the frequency distribution of the
expected number of detections becomes
\begin{equation}
 \frac{dN}{d \ln f} \approx 0.822 \times \frac{\pi^{1/3} G^{10/3}
  \mathcal{M}^{10/3} R}{3^{1/2} 5^{1/2} c^7 \rho_\mathrm{thr}^3}
  \frac{f^{-2/3} T^{3/2}}{[(3/20)S(f)]^{3/2}} \label{eq:mono}
\end{equation}
in the monochromatic approximation.

A remarkable point is that the expected number of detections is
proportional to $\mathcal{M}^{10/3}$. This expression implies that our
simple estimate should also be valid approximately for the case that the
chirp mass has distribution, once we replace $\mathcal{M}$ by the
averaged chirp mass defined by
\begin{equation}
 \langle \mathcal{M} \rangle \equiv \left[ \int p ( \mathcal{M} )
  \mathcal{M}^{10/3} d \mathcal{M} \right]^{3/10} , \label{eq:average}
\end{equation}
where $p ( \mathcal{M} )$ is the probability distribution of the chirp
mass. For example, if the realistic typical chirp mass would be $\sim
9M_\odot$ corresponding to a $10M_\odot$--$10M_\odot$ binary, the number
of detections is smaller by a factor of $\sim 40$--50 than the current
estimate. This dependence should be compared with $\mathcal{M}^{5/2}$
expected for ground-based detectors (see equations \ref{eq:snrav} and
\ref{eq:snrfac}), which observe chirping signals.

The distribution scales as $\propto T^{3/2}$ because of the relation
$\rho \propto T^{1/2}$. This means that a longer operation of eLISA will
detect more binary black holes than the increase linear in time
differently from compact binary coalescences for ground-based
detectors. The longer operation will further increase the detections of
non-merging binary black holes by removing more confusion noise caused
by Galactic compact binaries, and thus $T^{3/2}$ is conservative. While
this effect will not be relevant to merging binary black holes at high
frequency, their detections should also increase even faster than
$T^{3/2}$, because the longer operation of eLISA allows binaries at
lower frequency to merge within the observation period.

\begin{figure}
 \includegraphics[width=.95\linewidth]{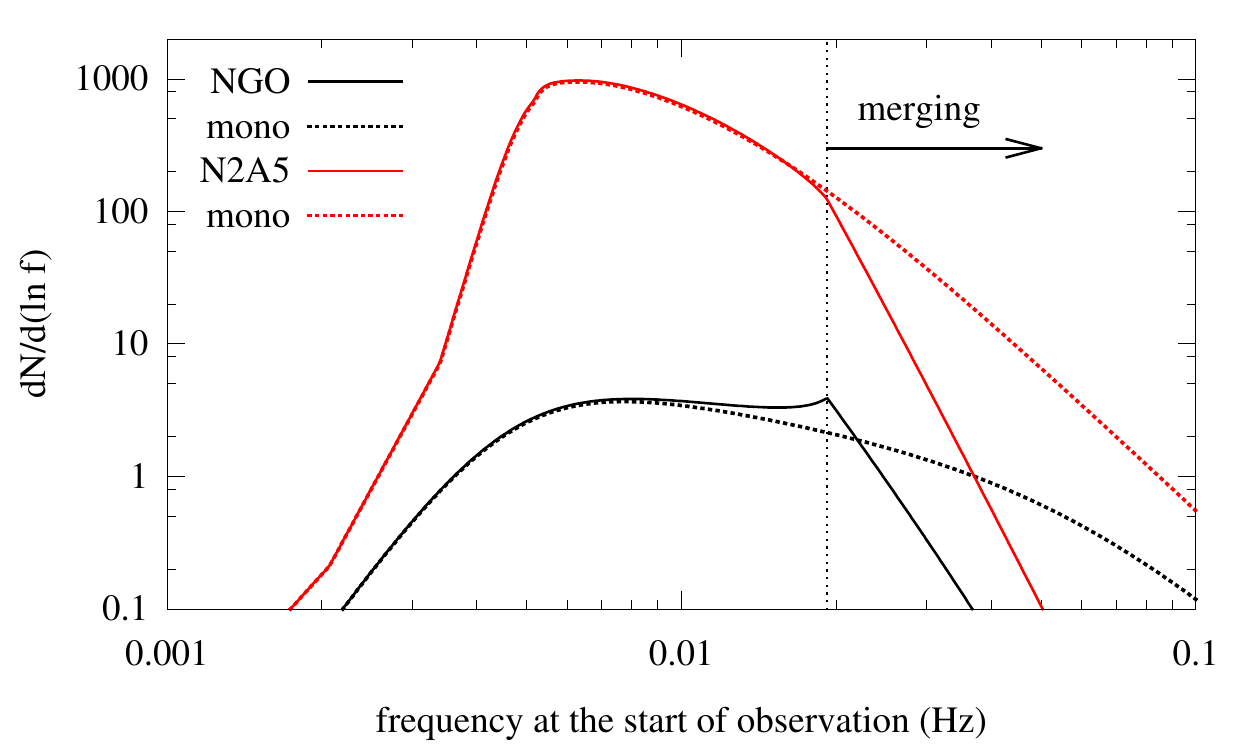} \caption{Comparison of
 the frequency distribution of the expected number of detections
 obtained by the exact expression, equation \eqref{eq:exact} (solid
 curve), and that by the monochromatic approximation, equation
 \eqref{eq:mono} (dashed curve). The results for NGO (black) and N2A5
 (red) configurations are shown assuming 3-yr observations. The vertical
 dotted line marks $f_\mathrm{merge} ( \SI{3}{yr} ) =
 \SI{19.2}{\milli\hertz}$.} \label{fig:mono}
\end{figure}

Fig.~\ref{fig:mono} compares $dN/d \ln f$ obtained by the exact
integration, equation \eqref{eq:exact}, and that by the monochromatic
approximation, equation \eqref{eq:mono}, for $T=\SI{3}{yr}$. The
agreement is quite good at frequency lower than $f_\mathrm{merge} (T)$,
particularly for N2A5. The deviation is only by a factor of less than 2
at $f_\mathrm{merge} (T)$ even for NGO. This clearly shows that the
monochromatic approximation works quite well to estimate the number of
non-merging binary black holes. By contrast, the monochromatic
approximation significantly overestimates the number of merging binary
black holes at high frequency. The breakdown of this approximation at $f
\gtrsim f_\mathrm{merge} (T)$ is inevitable, because the binary
evolution necessarily becomes important. Still, taking the fact that the
detections is dominated by non-merging binaries, the monochromatic
approximation is useful to derive semiquantitative dependence of
characteristic quantities on relevant parameters and to evaluate
detector performance by a concise calculation.

\begin{figure}
 \includegraphics[width=.95\linewidth]{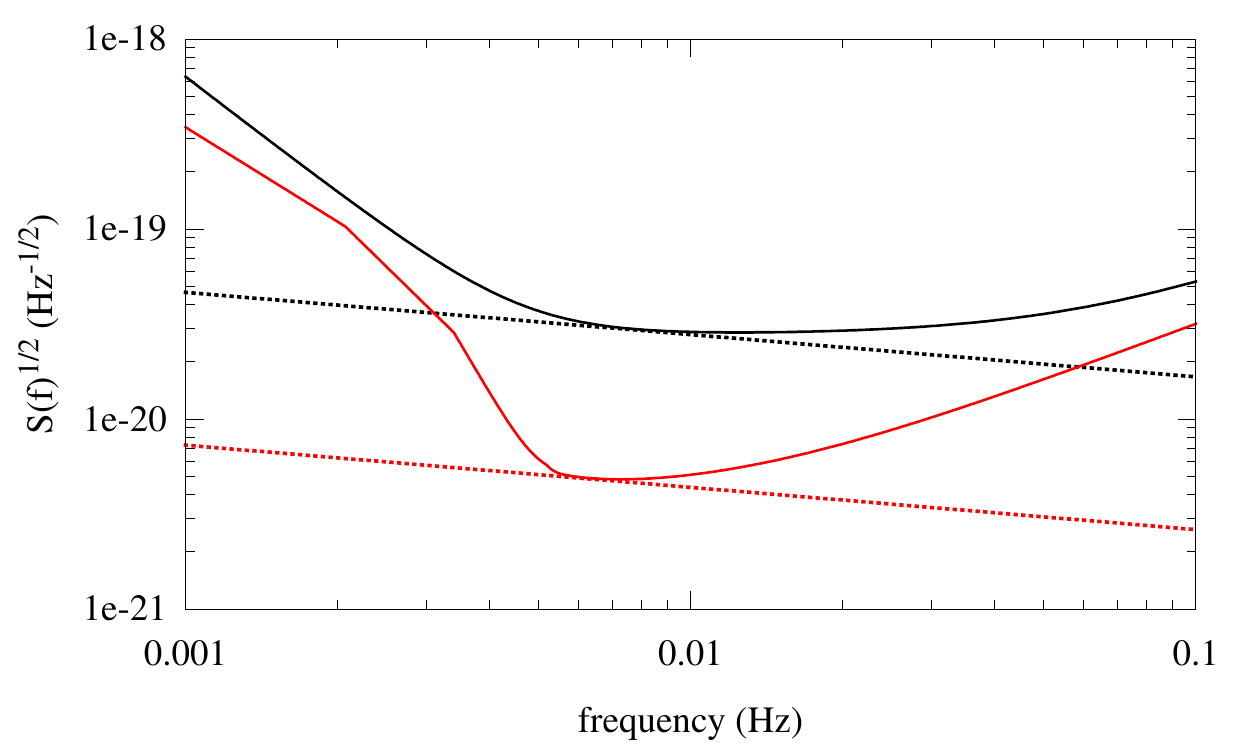} \caption{Tangential
 line with the slope $-2/9$ to the square root of $S(f)$. The black and
 red curves are for NGO and N2A5, respectively.} \label{fig:tangent}
\end{figure}

The frequency at which the distribution, $dN/d \ln f$, peaks will
dominate the detections, and it can be evaluated graphically by drawing
a tangential line with an appropriate slope to the noise spectral
density. Specifically, equation \eqref{eq:mono} shows that the peak
frequency is obtained as a contact point of a tangential line with the
slope $-2/9$, i.e., $\propto f^{-2/9}$, to the square root of the noise
spectral density, $S(f)^{1/2}$, as shown in Fig.~\ref{fig:tangent}. The
same can be done by taking the minimum of $S(f) f^{4/9}$.

\section{localization error}

We briefly estimate the localization error using empirical formulae
derived in \citet{takahashi_seto2002} adopting the monochromatic
approximation and Fisher analysis \citep[see
also][]{cutler1998,vecchio_wickham2004}. The errors of the sky location
and the distance to the source for $T \gtrsim \SI{2}{yr}$ are,
respectively, given by (recall $\SI{1}{deg^2} = \SI{3.e-4}{str}$)
\begin{align}
 \Delta \Omega (f) & \sim \SI{7.1e-4}{str} \left( \frac{\rho}{10}
 \right)^{-2} \left( \frac{f}{\SI{10}{\milli\hertz}} \right)^{-2} ,
 \label{eq:sky} \\
 \frac{\Delta D}{D} & \sim 0.2 \left( \frac{\rho}{10} \right)^{-1} .
\end{align}
Here, the empirical formula of $\Delta \Omega (f)$ is applicable to $f
\gtrsim \SI{2}{\milli\hertz}$ where the source is localized primarily by
the Doppler shift associated with the annual motion of the detector
\citep{takahashi_seto2002}. Thus, the formula is appropriate for the
frequency range that we are interested in this study (see
Fig.~\ref{fig:range}). Assuming the error volume $\Delta V$ to have an
elliptical shape in conformity with the Fisher analysis, we find
\begin{align}
 \Delta V (f) & \sim \frac{4}{3} D^2 \Delta \Omega (f) \Delta D \\
 & \sim \SI{200}{Mpc^3} \left( \frac{D}{\SI{100}{Mpc}} \right)^3 \left(
 \frac{\rho}{10} \right)^{-3} \left( \frac{f}{\SI{10}{\milli\hertz}}
 \right)^{-2} . \label{eq:errvol}
\end{align}
While this expression is valid for any monochromatic source, the
signal-to-noise ratio, $\rho$, is determined by $D$ and $f$ for binary
black holes with given binary parameters such as the chirp mass and the
inclination angle.

In this section, we focus on the nearest and thus loudest binaries
located at $D_\mathrm{near}(f)$, because the error volume should be the
smallest for such systems and the host galaxy should be determined most
accurately. To derive the signal-to-noise ratio, we average the
inclination angle somewhat arbitrarily in the same manner as is done in
the previous section, equation \eqref{eq:ave}. This averaging has a
useful feature that the signal-to-noise ratio for the nearest binary
becomes $\rho_\mathrm{thr} D_\mathrm{eff} / D_\mathrm{near}$. The error
volume for the nearest binary $( \Delta V )_\mathrm{near}$ is found to
be proportional to $D_\mathrm{near}^6 D_\mathrm{eff}^{-3} f^{-2}$
irrespective of the precise averaging procedure, and furthermore in the
monochromatic approximation, proportional to $f^{4/3} S(f)^{3/2}$. This
implies that the smallest error volume is achieved at the minimum of
$S(f) f^{8/9}$ or equivalently the contact point of $S(f)^{1/2}$ and a
tangential line with the slope $-4/9$. For eLISA noise curves, the
minimum of $( \Delta V )_\mathrm{near}$ is very close to the maximum of
$dN/d \ln f$.

\begin{figure}
 \includegraphics[width=.95\linewidth]{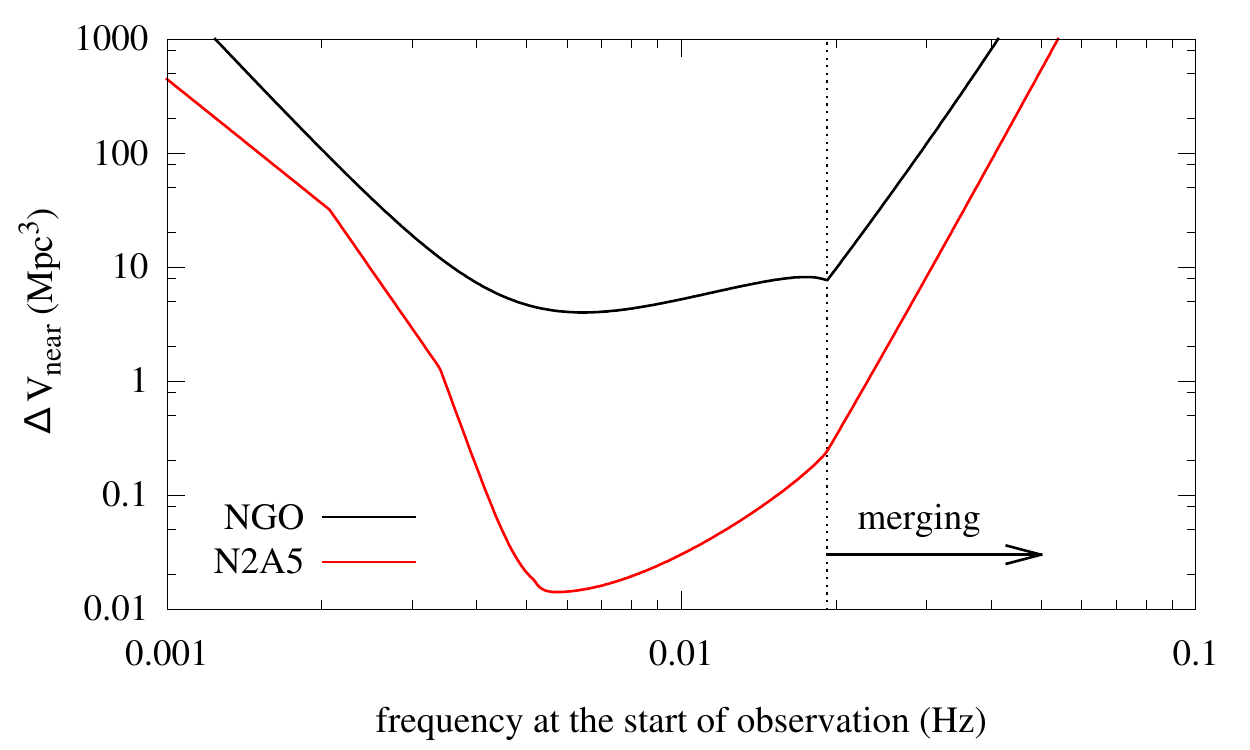} \caption{Error volume
 $\Delta V(f)$ for the nearest and thus loudest binaries located at
 $D_\mathrm{near} (f)$ for 3-yr observations of eLISA. The black and red
 curves are for NGO and N2A5, respectively. It should be cautioned that
 the result below $\sim \SI{2}{\milli\hertz}$ is poorly described by the
 empirical formula adopted in this study, equation \eqref{eq:sky}. The
 vertical dotted line marks $f_\mathrm{merge} ( \SI{3}{yr} ) =
 \SI{19.2}{\milli\hertz}$.} \label{fig:volume}
\end{figure}

Fig.~\ref{fig:volume} shows the error volume for the nearest and thus
loudest binaries in each logarithmic frequency interval. This figure is
drawn using the exact integration, equation \eqref{eq:exact}, rather
than the monochromatic approximation, equation \eqref{eq:mono}, and the
results are nearly identical below $f_\mathrm{merge}(T)$. Assuming the
typical number density of galaxies $\sim \SI{0.01}{Mpc^{-3}}$, this
figure strongly suggests that we may be able to determine the host
galaxy of many binary black holes with very high confidence. Because
massive binary black holes like GW150914 are hardly expected to be
localized accurately by ground-based detectors, which crucially rely on
high-frequency signals with $f \gtrsim \SI{100}{\hertz}$ via the
triangulation \citep{fairhurst2009}, the exquisite accuracy of
localization by eLISA will be invaluable to study the origin of massive
binary black holes. The accurate determination of the host galaxy will
also be important to study the cosmology using binary black holes as
standard sirens \citep{schutz1986}.

Our estimate refers only to the statistical error. Taking the high
precision of localization by eLISA, other sources of errors such as the
inaccuracy of templates would require serious consideration
\citep{cutler_vallisneri2007}. Furthermore, the error volume estimated
above is not accurate at high frequency, because the monochromatic
approximation used to derive equation \eqref{eq:errvol} breaks down. The
estimate for $f \lesssim \SI{2}{\milli\hertz}$ is also inaccurate,
because the empirical formula does not apply. While the statistical
error in the frequency range where most of the binaries are detected is
handled properly in our discussion, it would be interesting to study the
localization error in a more comprehensive manner along the line of our
concise calculation to complement Monte Carlo simulations
\citep{nishizawa_bks2016,sesana2016,vitale2016}. In addition, the
eccentricity estimation will also be important to clarify the nature of
massive binary black holes, and thus it would be useful to assess the
estimation error in a quantitative manner. The eccentricity may be
estimated via the amplitude of the third harmonic mode in the
monochromatic limit \citep{seto2016} or via the matched-filtering
analysis at high frequency, for which \citet{nishizawa_bks2016} gave
error distribution by Monte Carlo simulations combined with the Fisher
analysis. We left such extension for the future study.

\section{summary}

We investigate the prospects for detecting extragalactic stellar-mass
binary black holes by eLISA, motivated by the first direct detection of
gravitational waves, GW150914. eLISA might observe as many binary black
holes similar to GW150914 as supermassive binary black holes. The
detection will be dominated by binaries that do not merge within the
observation period $T$ of eLISA, particularly when the low-frequency
acceleration noise is improved. The total number of detections varies by
about two orders of magnitude depending on the detector sensitivity, and
a long-arm detector is highly desired for promoting multiband
gravitational-wave astronomy by increasing the detections of merging
binary black holes. Scientific returns from a long operation is more
than the increase linear in time. Specifically, the total number of
detections increases as or faster than $T^{3/2}$, and the number of
merging binary black holes increases much faster than $T^{3/2}$.

A monochromatic approximation works well to derive various features of
observational prospects, especially for the case that the sensitivity is
high at low frequency so that the detections of non-merging binary black
holes are numerous. The expected number of detections is found to be
proportional to $\mathcal{M}^\mathrm{10/3}$ by the monochromatic
approximation, and our estimate will be applicable to binary black holes
with chirp-mass distribution if we use the appropriately weighted
average chirp mass (equation \ref{eq:average}). The error volume of the
localization can be so small that the host galaxy is determined with
high confidence. This ability of eLISA will be invaluable to clarify the
origin of massive binary black holes.

\section*{Acknowledgements}

This work is supported by JSPS Kakenhi Grant-in-Aid for Research
Activity Start-up (No.~15H06857), for Scientific Research (No.~15K65075)
and for Scientific Research on Innovative Areas (No.~24103006).
Koutarou Kyutoku is supported by the RIKEN iTHES project.

\bibliographystyle{mn2e}
%\bibliography{paper}

\end{document}